\documentclass[prl,twocolumn,showpacs,showkeys,superscriptaddress,citeautoscript,footnote]{revtex4-1}
 
\usepackage{amsmath,amssymb,amsfonts,multirow,slashed}

\usepackage[colorlinks=true,citecolor=blue,linkcolor=blue]{hyperref}
\usepackage{physics}

\def\al#1{\alpha_{#1}}

\usepackage{acronym}

\usepackage{todonotes}

\usepackage{microtype}

\def\eq#1{{Eq.~(\ref{#1})}}

\def\Tr{\mbox{Tr}\,}

\newcommand\lsim{\mathrel{\rlap{\lower4pt\hbox{\hskip1pt$\sim$}}
    \raise1pt\hbox{$<$}}}
\newcommand\gsim{\mathrel{\rlap{\lower4pt\hbox{\hskip1pt$\sim$}}
    \raise1pt\hbox{$>$}}}

\newcommand{\beq}{\begin{equation}}
\newcommand{\eeq}{\end{equation}}
\newcommand{\bea}{\begin{eqnarray}}
\newcommand{\eea}{\end{eqnarray}}
\newcommand{\bem}{\begin{pmatrix}}
\newcommand{\eem}{\end{pmatrix}}
\newcommand{\noi}{\noindent}
\newcommand{\non}{\nonumber}

\newcommand{\bet}{\begin{itemize}}
\newcommand{\eet}{\end{itemize}}
\newcommand{\ben}{\begin{enumerate}}
\newcommand{\een}{\end{enumerate}}

\begin{document}

\title{Safe Hologram}

\author{Borut Bajc}
\email{borut.bajc@ijs.si}
\affiliation{J. Stefan Institute, 1000 Ljubljana, Slovenia}
 \author{Adri\'an Lugo}
\email{lugo@fisica.unlp.edu.ar}
\affiliation{Instituto de F\'isica de La Plata-CONICET, and Departamento de F\'isica,
Facultad de Ciencias Exactas, Universidad Nacional de La Plata, Argentina}
\author{Francesco Sannino}
\email{sannino@cp3.sdu.dk}
\affiliation{CP3-Origins \& the Danish Institute for Advanced Study. University of Southern Denmark. Campusvej 55, DK-5230 Odense, Denmark}

\begin{abstract}
We introduce a holographic model encapsulating the dynamics of safe quantum field theories. 
\end{abstract}
\preprint{}

\maketitle

\acrodef{vev}{vacuum expectation value}
\acrodef{cft}[\textsc{cft}]{conformal field theory}
\acrodefplural{cft}[\textsc{cft}s]{conformal field theories}
\acrodef{eft}[\textsc{eft}]{effective field theory}
\acrodef{nlsm}[\textsc{nlsm}]{non-linear sigma model}

\label{sec:intro}

The  discovery of a holographic description of ${\cal N} = 4 $ super Yang-Mills  led to a revolution  in our understanding of four dimensional gauge theories known as the AdS-CFT correspondence \cite{Maldacena:1997re,Gubser:1998bc,Witten:1998qj}.  See also \cite{Bianchi:2001de,Bianchi:2001kw,Girardello:1998pd,Girardello:1999hj,Girardello:1999bd,Freedman:1999gp} for an incomplete list of references useful for this work. For reviews on the subject see, for example, \cite{Aharony:1999ti,DeWolfe:2000xi,DHoker:2002nbb,Skenderis:2002wp,DeWolfe:2018dkl}. Going away from maximally supersymmetric theories the community  concentrated on capturing the non-perturbative dynamics of asymptotically free quantum field theories via holographic toy-models \cite{Erlich:2005qh,Csaki:2006ji,Gursoy:2007cb,Gursoy:2007er,Jarvinen:2011qe,Kiritsis:2016kog}.  One of the important properties of asymptotic freedom \cite{Gross:1973ju,Politzer:1973fx} is that it is a well defined theory at short distance. This is so because in the UV is controlled by a non-interacting CFT. Another way of  expressing the same point is that the theory is cutoff free. This should be contrasted with purely four dimensional scalar field theories that are trivial because the only way to eliminate the UV cutoff is for the theory to be non-interacting at all scales \cite{Luscher:1987ay,Luscher:1987ek, Luscher:1988uq}.  This is also known as the triviality problem, which limits the predictivity  to a scale of maximal  UV extension \cite{Callaway:1988ya}. 
Freedom is, however, not the only solution to this problem as long as the UV is controlled by a CFT which can be interacting. This  is the  safe scenario \cite{Weinberg:1980gg}.  

The notable difference between  freedom and  safety  is that canonical power counting  is modified at high energy. 
Couplings may become large in the UV and small expansion parameters are often unavailable complicating the task of discovering asymptotic safety. Nevertheless, by taking the space-time dimensionality as a continuous parameter 
a few rigorous results for asymptotically safe UV fixed points exist \cite{Gastmans:1977ad,Christensen:1978sc,Weinberg:1980gg,Peskin:1980ay,
Gawedzki:1985uq,Gawedzki:1985ed,Morris:2004mg,Codello:2016muj}  in the spirit of the $\epsilon$-expansion \cite{Wilson:1973jj}, or by using large-$N$ techniques  \cite{Tomboulis:1977jk,Tomboulis:1980bs,Smolin:1981rm,
deCalan:1991km,Kazakov:2007su,Antipin:2011ny,Antipin:2011aa,
Antipin:2012kc,Antipin:2013pya}.

The discovery of controllable purely four dimensional safe quantum field theories \cite{Litim:2014uca,Litim:2015iea} demonstrated that the class of fundamental quantum field theories \cite{Wilson:1971bg,Wilson:1971dh}  is much wider than previously envisioned.  Motivated by this we  engineer the first safe hologram by assuming that the UV and IR conformal dynamics to be described by AdS spaces. 

The physics of an UV safe fixed point that is linked to an IR gaussian one (related to the loss of asymptotic freedom) can be captured by this  schematic form of the beta function 
\begin{equation}
\label{beta}
\beta(A) = b A^2(A_{UV} - A) \quad {\rm with} \quad  b>0 \ , 
\end{equation}
 with $A = g^2/(4 \pi)^2$  the gauge coupling strength. %
%
%
%

The simplified five-dimensional bulk theory that we take to describe the gauge dynamics reads 
\begin{equation}
\label{action}
S_{bulk}=\int d^dxdz\sqrt{|det(g_{ab})|}\left[\frac{1}{2}g^{ab}\left(\partial_a\chi \partial_b\chi \right)+
V (\chi)   \right] \ .
\end{equation}
\noi
Here we do not consider the global flavour dynamics that can be added later and assume the following non-dynamical metric (i.e. no back-reaction, $M_{Pl}\to\infty$ limit \cite{Bajc:2012vk,Bajc:2013bza}) 
\beq
ds^2=g_{ab}dx^adx^b=\frac{L^2}{z^2}\left(dz^2+dx_d^2\right) \ . 
\eeq
$\chi$ is the dilaton and its bulk profile is taken to reproduce the running of the 4D coupling constant via the relation

\beq
e^{\chi(z)}= A(z) \ ,
\label{chiofA}
\eeq
\noi
with the following relation between the bulk coordinate $z$ and the RG scale $\mu$
\beq
z=\frac{1}{\mu} \ .
\eeq
  In the following, thanks to \eqref{chiofA}, we will write all the relevant functions directly in terms of $A$. 

It is worth recalling that in the safe scenario the limit $z=0$ corresponds to the UV and it describes an interacting CFT.  At  $z=\infty$ we are in the deep IR and therefore we are describing a non-interacting CFT. 
  
  \bigskip
It is convenient to introduce a superpotential-like object $W(A)$ related to the dilaton potential via 
\begin{equation}
L^2V(A) = \frac{1}{2}  \left(A\partial_A W(A)\right)^2 - d\, W(A) \ . 
\end{equation}
 with $d $ the dimension of the boundary space time. We are assuming no gravitational back reaction. This is the reason why we have  a term linear rather than quadratic in $W$ \cite{Bajc:2013wha}. 
 
 The equation of motion of $A(z)$ reads: 
 \begin{equation}
  z\partial_z A = A^2 \partial_A W(A) \ .
 \end{equation}
 By definition the above is minus the beta function of $A$ given in \eqref{beta}. This yields: 
 \begin{equation}
 \label{AW}
\partial_AW = - \frac{\beta(A)}{A^2} \ .
 \end{equation}
To obtain simple analytic solutions it is convenient to split the double zero at the origin of the beta function into two single zero one at the origin and the other at a distance $\epsilon$ from the origin. The second is an IR interacting Wilson-Fisher fixed point of the theory in $d=4-\epsilon$ dimensions. We shall denote this fixed point with $A_{IR} = \epsilon/(bA_{UV})$ and re-write the beta function accordingly. 
\begin{equation}
\label{betaepsilon}
\beta(A) = b A(A-A_{IR})(A_{UV} - A ) \ , 
\end{equation}
 up to $\epsilon$ corrections to the original $A_{UV}$. The above returns \eqref{beta} in the $\epsilon = 0 $ limit.  Working with \eqref{betaepsilon} we have:
 \begin{equation}
 W(A) = b \left[ \frac{1}{2} A \left( A - 2(A_{IR} + A_{UV}) \right) + A_{UV}A_{IR} \log\left(A\right)\right] \ , 
 \end{equation}
 up to an integration constant. 
 
 One of the important quantities characterising the specific CFT is the scaling exponents related to the first derivative of the beta function at a fixed point. These are also related to the operator dimensions of the theory  field strength   $F_{\mu \nu}F^{\mu \nu}$.  
 The scaling exponents can be readily computed from the beta function via \cite{Nielsen:1977sy,Spiridonov:1984br,Grinstein:1988wz,Gubser:2008yx,NunesdaSilva:2016jfy,Cata:2018wzl}
 \begin{equation}
 \label{def-D}
 \Delta_{F^2} = d + \partial_A \beta -  \frac{\beta}{A} \ .
 \end{equation}
A factor of two appearing in front of ${\beta}/{A}$ in \cite{Gubser:2008yx} is incorrect \cite{NunesdaSilva:2016jfy,Cata:2018wzl}. 
The equation above yields at the two  non-trivial fixed points  
\begin{eqnarray}
\label{IRUV}
\Delta_{IR} & =& d + b A_{IR}\left(A_{UV} - A_{IR} \right) \ , \nonumber \\  
\Delta_{UV} & =&  d - b A_{UV}\left(A_{UV} - A_{IR} \right) \ .
\end{eqnarray} 
In the holographic picture we have 
 \begin{equation}
\partial_{\log A}^2 (V) = \Delta \left(\Delta - d \right)   \ ,
 \end{equation}
 evaluated at the $UV$ and $IR$ fixed points. One finds either 
 \begin{equation}
 \label{Deltas}
 \Delta = \partial_{\log A}^2 W \quad {\rm or } \quad  d- \Delta = \partial_{\log A}^2 W \ ,
 \end{equation}
 ad the fixed points. Consistency with \eqref{def-D} requires to use the second expression so that we  reproduce \eqref{IRUV}.  The correct anomalous dimension of the field strength, i.e. without the factor of 2 in front of $\beta/A$, can be independently derived from the holographic construction when substituting  \eqref{AW} in the second relation of \eqref{Deltas}. 
 
 We are now ready to solve for the background field in the coupling range $A_{IR} \leq A \leq A_{UV}$ by integrating the beta function 
 \begin{equation}
 \label{zofA}
 z[A] = \exp\left(- \int \frac{dA}{\beta(A)}\right) \ .
 \end{equation}  
 The solution reads
 \begin{equation}
 \label{solzofA}
\left( \frac{z}{z_0}\right)^{bA_{IR}A_{UV}(A_{UV}- A_{IR})} = A^{A_{UV}-A_{IR}} \frac{(A_{UV} - A)^{A_{IR}}} 
 {(A-A_{IR})^{A_{UV}}} \ .
  \end{equation}
Via \eqref{chiofA} we have also the bulk profile for the dilaton $\chi(z)$. To compute relevant correlation functions in d dimensions we now solve the equation of motion for $\chi(z,x)$ 
\begin{equation} 
\left(z^2 \partial^2_z + \partial_x^2\right)\chi(z,x) - (d-1)z\partial_z \chi(z,x) = \partial_\chi V(\chi) \ .
\end{equation}
Fourier-transforming 
\begin{equation}
\chi(z,x) = \chi(z) + \int\frac{d^d k}{(2\pi)^d} e^{ikx} \xi(z,k) \ , 
 \end{equation}
and linearizing in $\xi$  we get
\begin{equation}
 z^2 \partial^2_z \xi(z,k) - (d-1)z\partial_z \xi(z,k) = \left[ k^2 z^2 + \partial_\chi^2 V(\chi(z)) \right] \xi(z,k) \ .\\ 
\end{equation} 
According to the holographic dictionary, knowing $\xi(z,k)$ allows us to determine the two point correlation function 
\begin{equation} 
G_2(k) =  \int \frac{d^d x}{(2\pi)^d} e^{-ikx}\langle F^2(x) F^2(0) \rangle  \ , 
\end{equation}
 from the $z\rightarrow 0$ limit through the identification
 \begin{equation}
 \label{G2}
 \xi(z,k) \rightarrow \xi_0(k)  \left(z^{d-\Delta_{UV}} + G_2(k)z^{\Delta_{UV}} \right)   \ .
 \end{equation}
 We now test the correctness of the holographic dual by derermining an analytic solution for $G_2$ in the $k\rightarrow 0$ limit.

  In general it is not possible to obtain an exact analytic expression for $\xi(z,k)$. Therefore we will 
  use an approximate method - the matching procedure \cite{Hoyos:2012xc,Bajc:2013wha,Hoyos:2013gma}: we 
  first compute the solution in the $z\rightarrow \infty$ limit. Here the equation simplifies to
  \begin{eqnarray}
 \left[z^2 \partial^2_z   - (d-1)z\partial_z  -  k^2 z^2  -  \Delta_{IR}\left(\Delta_{IR} - d \right) \right] \xi(z,k)  = 0\ ,  \nonumber \\
  \end{eqnarray} 
 and with the well behaved solution for any $k$ and large $z$ we have  
 \begin{equation}
 \label{inftyxi}
 \xi_{\infty}(z,k) = 2 \left(\frac{kz}{2}\right)^{d/2} K_{\nu_{IR}} (kz) \quad {\rm with} \quad \nu_{IR} = \Delta_{IR} - \frac{d}{2} \ .
 \end{equation}
 Another useful limit is for $k\rightarrow 0$ and arbitrary $z$ for which we have 
 \begin{equation}
 z^2 \partial^2_z \xi(z,k) - (d-1)z\partial_z \xi(z,k) =  \partial_\chi^2 V(\chi(z))   \xi(z,k) \ .\\ 
\end{equation} 
The solution in this limit can be written as 
\begin{equation}
\xi_{k\rightarrow 0}(z,k) = c_1(k) \xi_1(z) + c_2(k) \xi_2(z) \ , 
\end{equation}
 where 
\begin{eqnarray}
\xi_1(z) &=&  \frac{\partial_{\log z} A}{A}  \ ,   \\
\xi_2(z) &=&  \xi_1 (z) \int dz \frac{z^{d-1}}{\xi_1^2 (z)} .  
\end{eqnarray}
In fact, it is convenient to express $z$ as function of $A$ using \eqref{zofA} that leads to: 
\begin{eqnarray}
\xi_1(z[A]) &=&  - \frac{\beta(A)}{A}  \ , \\
\xi_2(z[A]) &=&  - \xi_1 (z[A]) \int \frac{dA}{\beta(A)^3} A^2{z^{d}[A]} .  \end{eqnarray}
The computation yields: 
\begin{align}
\xi_1&=-b(A-A_{IR})(A_{UV}-A) \ , \\
\xi_2/\xi_1
&=\left(\frac{z_0^d}{b^3}\right)\frac{\left(A_{UV}-A_{IR}\right)^{\alpha_{1}}\left(A_{UV}\right)^{\alpha_3}\left(A_{UV}-A\right)^{\alpha_2+1}}{\alpha_2+1} \nonumber  \\&
\times F_1\left(\alpha_2+1;-\alpha_1,-\alpha_3;\alpha_2+2;\frac{A_{UV}-A}{A_{UV}-A_{IR}},\frac{A_{UV}-A}{A_{UV}}\right) \ ,\non
\end{align}
with 
\begin{eqnarray}
\alpha_1&=& \frac{d}{d-\Delta_{IR}}-3 \ ,\\
\alpha_2&=& \frac{d}{d-\Delta_{UV}}-3 \ ,\\
\alpha_3&=& - \frac{d}{d-\Delta_{IR}}-\frac{d}{d-\Delta_{UV}}-1 \ ,
\end{eqnarray}
while the Appell  function $F_1$ is defined via the series
\begin{equation}
F_1(a;b,c;d;x,y)=\sum_{m=0}^{\infty}\sum_{n=0}^{\infty}\frac{(a)_{m+n}(b)_m(c)_n}{(d)_{m+n}\,m!\,n!}x^my^n \ ,
\end{equation}
with
\begin{equation}
(q)_k=\frac{\Gamma(q+k)}{\Gamma(q)} \ ,
\end{equation}
the Pochhammer symbol. The Appell series is well defined for $|x|, |y|<1$, which is the region we are interested in.

To determine the ratio $c_1/c_2$ we consider the $z\rightarrow \infty$ limit of $\xi_1$ and  $\xi_2$ and compare it with the $k\rightarrow 0$ limit of $\xi_{\infty}$ in \eqref{inftyxi}. In this limit we have: 
 \begin{equation}
 \label{inftyxi2}
 \lim_{k\rightarrow 0}\xi_{\infty}(z,k) =  \left(\frac{kz}{2} \right)^{
 \Delta_{IR}} \Gamma\left( - \nu_{IR}\right) + \left(\frac{kz}{2} \right)^{d-\Delta_{IR}} \Gamma\left( \nu_{IR}\right)    \ .
 \end{equation}
Expanding  $\xi_i$ with $i=1,2$ around the $z\rightarrow \infty$ limit we have 
\begin{eqnarray}
{\xi_1}_{z \rightarrow \infty}(z)  &=&  \widetilde{a}_{\infty}  \left(\frac{z}{z_0}\right)^{d-\Delta_{IR}}  \ , \nonumber \\ 
{\xi_2}_{z \rightarrow \infty}(z)  &=&{b}_{\infty}  \left(\frac{z}{z_0}\right)^{\Delta_{IR}} + \widetilde{b}_{\infty}  \left(\frac{z}{z_0}\right)^{d-\Delta_{IR}}    \ , \nonumber \\ 
\end{eqnarray}
with
\begin{align}
\tilde a_{\infty}&=-bA_{IR}^{\frac{2d-\Delta_{IR}-\Delta_{UV}}{d-\Delta_{UV}}}
\left(A_{UV}-A_{IR}\right)^{\frac{\Delta_{IR}-\Delta_{UV}}{d-\Delta_{UV}}}\\
b_{\infty}&=\left(\frac{z_0^d}{b^2}\right)\left(\frac{\Delta_{IR}-d}{2\Delta_{IR}-d}\right)\non\\
&\times A_{IR}^{-2-\frac{d-\Delta_{IR}}{d-\Delta_{UV}}}
(A_{UV}-A_{IR})^{-2+\frac{d-\Delta_{IR}}{d-\Delta_{UV}}}\\
\tilde b_{\infty}&=
\frac{\left(\frac{z_0^d}{b^2}\right)\left(\frac{\Delta_{UV}-d}{2\Delta_{UV}-d}\right)\Gamma(\alpha_1+1)\Gamma(\alpha_2+2)}
{\Gamma(\alpha_1+\alpha_2+2)}\non\\
&\times\left(A_{UV}\right)^{-1-\frac{d}{d-\Delta_{IR}}-\frac{d}{d-\Delta_{UV}}}\left(A_{IR}\right)^{1+\frac{d-\Delta_{IR}}{d-\Delta_{UV}}}\non\\
&\times\left(A_{UV}-A_{IR}\right)^{-4+\frac{d}{d-\Delta_{IR}}+\frac{\Delta_{IR}}{d-\Delta_{UV}}}\non\\
&\times\;_2F_1\left(\alpha_2+1,-\alpha_3;\alpha_1+\alpha_2+2;\frac{A_{UV}-A_{IR}}{A_{UV}}\right) \ .
 \end{align} 
Comparing with \eqref{inftyxi2} we derive
\begin{eqnarray} 
c_1 & = & \frac{\Gamma(\Delta_{IR}-d/2)}{\widetilde{a}_{\infty}} \left( \frac{kz_0}{2}\right)^{d-\Delta_{IR}} \non\\
&-& \frac{\widetilde{b}_{\infty}\Gamma(d/2-\Delta_{IR})}{\widetilde{a}_{\infty}{b}_{\infty}} \left( \frac{kz_0}{2}\right)^{\Delta_{IR}}  \ ,\nonumber \\
c_2 & = &   \frac{\Gamma(d/2-\Delta_{IR})}{{b}_{\infty}} \left( \frac{kz_0}{2}\right)^{\Delta_{IR}} \ .
\end{eqnarray}
Around $z\rightarrow 0$ we can re-expand $\xi_i$ with $i=1,2$ as 
\begin{eqnarray}
\xi_1(z) &=&  \widetilde{a}_0 \, z^{d - \Delta_{UV}}    \ , \nonumber \\
\xi_2(z) &=& b_0 \, z^{\Delta_{UV}} + \widetilde{b}_0  \, z^{d - \Delta_{UV}} .  
\end{eqnarray} 
The information above allows us to determine, using \eqref{zofA}, the numerical coefficients of the leading $z^{\Delta_{UV}}$ and $z^{d-\Delta_{UV}}$ terms near $z\rightarrow 0$. 
\begin{align}
\tilde a_0&=-b(A_{UV}-A_{IR})^{\frac{\Delta_{UV}-\Delta_{IR}}{d-\Delta_{IR}}}
A_{UV}^{\frac{2d-\Delta_{IR}-\Delta_{UV}}{d-\Delta_{IR}}} \ ,\\
b_0&=\left(\frac{z_0^d}{b^2}\right)\left(\frac{\Delta_{UV}-d}{2\Delta_{UV}-d}\right)\non\\
&\times(A_{UV}-A_{IR})^{\frac{d-\Delta_{UV}}{d-\Delta_{IR}}-2}
(A_{UV})^{-\frac{d-\Delta_{UV}}{d-\Delta_{IR}}-2} \ ,\\
\tilde b_0&=0 \ .
 \end{align} 
We are finally able to derive $G_2(k)$ analytically in \eqref{G2} by expanding  $\xi_{k\rightarrow 0}$ around $z\rightarrow 0$ 
\begin{equation}
\label{G2computed}
G_2(k) = \frac{b_0}{\widetilde{a}_0}
\frac{c_2(k)}{c_1(k)} \propto \left(kz_0\right)^{2\Delta_{IR}-d}+\ldots\ .
\end{equation}
This result confirms the expected IR behaviour of the $F^2$  operator two point correlation function. It therefore demonstrates that our holographic engineering of a safe quantum field theory is well defined. One can now go beyond this initial investigation by determining the full $G_2$ dependence on $k$ either numerically or via an approximate analytic solution \cite{Bajc:2012vk}. 

So far we concentrated on the gauge degrees of freedom, however, in the future, one can also explore flavour symmetries by opportunely introducing new holographic bulk fields corresponding, for example, to a fermion bilinear at the  boundary.

 We are now ready to discuss the holographic engineering of the safe theory discovered in \cite{Litim:2014uca} which is an $SU(N_c)$ gauge theory featuring gauge fields $A_{\mu}^{a}$ with associated field strength $F_{\mu\nu}^{a}$, $N_f$ 
 Dirac  fermions $Q_i$ $(i=1, \cdots , N_f)$ transforming according to the fundamental representation of the gauge group (color-index muted), and an $N_f \times N_f$
complex matrix scalar field $H$ uncharged under the gauge group.
The fundamental action is  \cite{Antipin:2013pya,Litim:2014uca}: %
\bea
\label{F2}
L_{\rm YM}&=& - \tfrac{1}{2} \Tr \,F^{\mu \nu} F_{\mu \nu}\nonumber\\
\label{F}
L_F\ &=& \ \Tr\left(
\overline{Q}\,  i\slashed{D}\, Q \right)
\nonumber\\
\label{Y}
L_Y \ &=&
\ y \, \Tr\left(\overline{Q}_L H Q_R + \overline{Q}_R H^\dagger Q_L\right)
\nonumber\\
\label{H}
L_H \ &=& \ \Tr\,(\partial_\mu H ^\dagger\, \partial^\mu H) \nonumber\\
\label{U}
L_U \ &=&
-u\,\Tr\,(H ^\dagger H )^2  \,\nonumber\\
\label{V}
L_V \ &=&
-v\,(\Tr\,H ^\dagger H )^2  \,. 
\eea
$\Tr$ is the trace over both color and flavor indices, and the decomposition $Q=Q_L+Q_R$ with $Q_{L/R}=\frac 12(1\pm \gamma_5)Q$ is understood. 
In four dimensions, the model has four classically marginal coupling constants given by the gauge coupling $g$, the Yukawa coupling $y$, the quartic scalar coupling $u$ and the `double-trace' scalar coupling $v$, which we write as:
\begin{align}
	\label{couplings}
	\al g & =\frac{g^2\,N_c}{(4\pi)^{d/2}\Gamma(d/2)}\,,\quad
	\al y=\frac{y^{2}\,N_c}{(4\pi)^{d/2}\Gamma(d/2)}\,,\quad\nonumber\\
	\al h & =\frac{{u}\,N_f}{(4\pi)^{d/2}\Gamma(d/2)}\,,\quad
	\al v=\frac{{v}\,N^2_f}{(4\pi)^{d/2}\Gamma(d/2)}\,.
\end{align}
We have normalized the couplings with the appropriate powers of $N_c$ and $N_f$ so that in the Veneziano-Witten limit $N_f/N_c$ is a finite  number. In dimensions different from four the couplings are dimensionful. We will denote by 
$\beta_i$ with $i=(g,y,h,v)$ the $\beta$-functions for the dimensionless version of the couplings in \eq{couplings}, which we will still call $\alpha_i$ for simplicity. {  We express the $\beta$ functions in terms of  
\begin{equation}\delta= \frac{N_f}{N_c}-\frac{11}{2} \ ,\end{equation} 
which in the  Veneziano-Witten limit is a continuous parameter taking values in the interval $[-\tfrac{11}{2}, \infty)$.}

 In four dimensions, once asymptotic freedom is lost for $N_f> {11N_c}/{2}$, the theory develops an UV interacting fixed point controlled by  $1\gg \delta > 0$. which is connected to an infrared free gaussian fixed point. The UV critical surface is one-dimensional meaning that the Yukawa, and the two scalar couplings can be written as functions of the gauge coupling \cite{Litim:2014uca}. This also means that the UV and IR fixed points are on a globally defined RG line that separates different physical phases, the {\it separatrix}. On the (peturbative) separatrix  the two-loops {\it effective} gauge beta function can be written only in terms of the gauge coupling: 
\begin{align}\label{EffectiveBetaNLO}
\beta_g^ {\rm eff}=&-\epsilon\alpha_g+\left\{\frac{4\delta}{3}+\frac{2\epsilon\left(\delta+\tfrac{11}{2}\right)^2}{2\delta+13}\right\}\alpha_g^2\nonumber\\
\qquad&+\frac{2\left(8\delta^2+46\delta-57\right)}{3(2\delta+13)}\alpha_g^3\,.
\end{align}
The term linear in  $\alpha_g$ emerges in $d=4-\epsilon$ dimensions \cite{Codello:2016muj}. The phase diagram in $d=4-\epsilon$ dimensions has been studied in \cite{Codello:2016muj}. 
It is clear that the beta function of \eqref{EffectiveBetaNLO} has the same form of \eqref{betaepsilon} and with the identification $A = \alpha_g$ we have: 
\begin{eqnarray}
b & = &\frac{38}{13}-\frac{1424}{507} \delta  \ , \nonumber \\ 
A_{IR} & = & \frac{3}{4} \frac{\epsilon}{\delta}   \ , \nonumber \\ 
A_{UV} & = &  \frac{26}{57} \delta\ , 
\end{eqnarray}
to first order in $\epsilon$ and $\delta$ and in the limit $\epsilon \ll \delta \ll 1$.
We can now determine all the relevant holographic information in terms of the gauge-theory quantities such as: 
\begin{eqnarray}
\Delta_{IR} & = & 4 - \frac{171}{104} \frac{\epsilon^2}{\delta^2} \ , \nonumber \\ 
\Delta_{UV} & = & 4 - \frac{104}{171} \delta^2 \ . 
\end{eqnarray}
The above allows us to map the theory into its geometric description and compute the $F^2$ two-point function via \eqref{G2computed}. 
 
To determine correlators to flavour symmetries we  need to introduce the related bulk fields and re-determine their backgrounds and perturbations.  

\bigskip
We engineered the first holographic construction of safe quantum field theories. We have shown that the holographic dual permits to compute relevant correlators of the underlying gauge dynamics. We then applied our construction to the first known model of four dimensional safe quantum field theories.  In the future we aim at making the correspondence crispier by, for example, investigating the fate and impact of the global symmetries of the theory. 
The holographic engineering can also be used as a model of    non-perturbative safe dynamics emerging at finite number of flavours and colors in \cite{Litim:2014uca}. The formalism can also be employed to analyse the safe dynamics \cite{Pica:2010xq,Antipin:2017ebo} of non-abelian gauge-fermion theories at finite number of colors and very large number of flavours. Here one can use the knowledge of the   leading order in the 
$1/N_f$ beta function and fermion mass anomalous dimensions \cite{PalanquesMestre:1983zy,Gracey:1996he,Holdom:2010qs,Pica:2010xq}. Because  the UV zero in the beta function stems from a logarithmic singularity, consistency checks  are important~\cite{Antipin:2017ebo,Ryttov:2019aux}.  First principle lattice investigations have recently appeared  \cite{Leino:2019qwk} while early studies can be found in \cite{deForcrand:2012vh}. Finally, adding large charges in the holographic framework can be used to cross-check the construction against the large charge expansion for safe quantum field theories introduced in \cite{Orlando:2019hte}.

 \bigskip

\acknowledgments
The work of F.S. is partially supported by the Danish National Research Foundation under grant DNRF:90. B.B. and F.S. thank respectively CP3-Origins and CERN, theoretical physics department for the hospitality when completing  this work. B.B. acknowledges the financial support from the Slovenian Research Agency (research core funding No. P1-0035)

\end{document}